\documentclass[12pt]{article}
\usepackage{amssymb}

\newcommand{\beq}{\begin{equation}}
\newcommand{\eeq}{\end{equation}}
\newcommand{\beqn}{\begin{eqnarray}}
\newcommand{\eeqn}{\end{eqnarray}}

\newcommand{\B}{\mbox{${\bf B}$}}
\newcommand{\E}{\mbox{${\bf E}$}}

\newcommand{\p}{\mbox{${\bf p}$}}

\newcommand{\s}{\mbox{${\bf s}$}}
\newcommand{\x}{\mbox{${\bf x}$}}

\newcommand{\br}{\mbox{${\bf r}$}}

\newcommand{\bv}{\mbox{${\bf v}$}}
\newcommand{\bx}{\mbox{${\bf x}$}}

\newcommand{\Na}{\mbox{{\boldmath$\nabla$}}}

\newcommand{\val}{\mbox{{\boldmath$\alpha$}}}

\newcommand{\ep}{\mbox{${\varepsilon}$}}
\newcommand{\om}{\mbox{${\omega}$}}

\begin{document}

\begin{center}
{\bf \large Equations of Motion of Spinning Relativistic
Particles}\\

\bigskip

I.B. Khriplovich\footnote{khriplovich@inp.nsk.su}\\

\bigskip

Budker Institute of Nuclear Physics, 630090 Novosibirsk, Russia

\end{center}

\bigskip
\bigskip

\begin{abstract}
The motion of spinning relativistic particles in external
electromagnetic and gravitational fields is considered. Covariant
equations for this motion are demonstrated to possess pathological
solutions, when treated nonperturbatively in spin. A
self-consistent approach to the problem is formulated, based on
the noncovariant description of spin and on the usual,
``na\"{\i}ve'' definition of the coordinate of a relativistic
particle. A simple description of the gravitational interaction of
first order in spin, is pointed out for a relativistic particle.
The approach developed allows one to consider effects of higher
order in spin. Explicit expression for the second-order
Hamiltonian is presented. We discuss the gravimagnetic moment,
which is a special spin effect in general relativity.
\end{abstract}

\bigskip
\bigskip

\section{Introduction}

The problem of the motion of a particle with internal angular
momentum (spin) in an external field consists of two parts: the
description of the spin precession and accounting for the spin
influence on the trajectory of motion. To lowest nonvanishing
order in $c^{-2}$ the complete solution for the case of an
external electromagnetic field was given more than 70 years
ago~[1]. The gyroscope precession in a centrally symmetric
gravitational field had been considered to the same approximation
even earlier~[2]. The fully relativistic problem of the spin
precession in an external electromagnetic field was also solved
more than 70 years ago~\cite{fr}, and then in a more convenient
formalism, using the covariant vector of spin, in~\cite{bmt}.

The situation is different with the second part of the problem,
which refers to the spin influence on the trajectory. Covariant
equations of motion for a relativistic spinning particle in an
electromagnetic field were written in the same paper~\cite{fr},
and for the case of a gravitational field in~\cite{pa}. Then these
equations were discussed repeatedly from various points of view in
numerous papers. The problem of the influence of the spin on the
trajectory of a particle in external fields is not only of purely
theoretical interest. It is related to the description of the
motion of relativistic particles in accelerators~\cite{dk} (see
also review~\cite{hei}). There are also macroscopic objects for
which internal rotation certainly influences their trajectories.
We mean the motion of Kerr black holes in external gravitational
fields.

Here we will elucidate serious shortcomings of the covariant
description of the spin influence on the trajectory, and
formulate a self-consistent noncovariant approach to the problem.
Most of the results presented below were worked out in
collaboration with A. Pomeransky and R. Sen'kov. Our papers~[8-10]
contain more details, as well as more complete list of references.

\section{What is wrong with covariant\\ equations of motion?}

A covariant correction $f^{\mu}$ to the Lorentz force
$eF^{\mu\nu}u_{\nu}$ should be linear in the tensor of spin
$S_{\mu\nu}$ and in the gradient of the tensor of electromagnetic
field $F_{\mu\nu,\lambda}\,$, it may depend also on the 4-velocity
$u^\mu$. Since $u^{\mu}u_{\mu}=1$, this correction must satisfy
the condition $u_{\mu}f^{\mu}=0$. From the mentioned tensors one
can construct only two independent structures meeting the last
condition. The first, $
\eta^{\mu\kappa}F_{\nu\lambda,\kappa}S^{\nu\lambda}\,-\,
F_{\lambda\nu,\kappa}u^{\kappa}S^{\lambda\nu}u^{\mu}, $ reduces
in the $c^{-2}$ approximation to
\[
2\s(\B,_m-\,[\bv \times
\E,_m]),
\]
and the second, $
u^{\lambda}F_{\lambda\nu,\kappa}u^{\kappa}S^{\nu\mu}, $ reduces to
\[
\frac{d}{dt}\,[\s \times \E]\,.
\]
Here $\B$ and $\E$ are external magnetic and electric fields;
$e$, $m$, $\s$, and $\p$ are the particle charge, mass, spin, and
momentum, respectively; $g$ is its gyromagnetic ratio; a comma
with a subscript denotes a partial derivative. Obviously, no
linear combination of these two structures can reproduce the
correct result for the spin-dependent force,
\[
f_m =\,\,\frac{eg}{2m}\s\B,_m+\,\frac{e(g-1)}{2m}
\,\left(\frac{d}{dt}[\E \times \s\,]_m-\, \s [\bv \times
\E,_m\,]\right),
\]
which follows in the $c^{-2}$ approximation from the well-known
noncovariant Hamiltonian (see, e.g.,~\cite{blp})
\[ H_{e1} = {\bf \Omega}\s=\,\frac{e}{2m}\,\s\,
\left\{(g-2)\,\left[\B -\,\frac{\gamma}{\gamma+1}\,\bv(\bv\B)\,
-\,\bv \times \E \right]\,\right. \]
\begin{equation}\label{lse}
\left.+2\,\left[\frac{1}{\gamma}\,\B -\,\frac{1}{\gamma+1}\,\bv
\times \E\right]\right\},\;\;\; \gamma =\,{1 \over
\sqrt{1-v^2}}\,.
\end{equation}
Let us emphasize, that though being noncovariant, this
Hamiltonian is valid for arbitrary velocities.

Here the covariant formalism can be reconciled with the correct
results if the coordinate $\x$ entering the covariant equation is
related to the usual one $\br$ as follows in the $c^{-2}$
approximation:
\begin{equation}\label{def}
\x\,=\,\br+\,\frac{1}{2 m}\s\times\bv.
\end{equation}
The generalization of this substitution to the case of arbitrary
velocities is~\cite{hei}:
\begin{equation}\label{defh}
\x\,=\,\br+\,\frac{\gamma}{m(\gamma+1)}\s\times\bv.
\end{equation}

However, after this velocity-dependent substitution, the equations
of motion depend on the third time derivative of coordinate. This
is harmless by itself as long as the corresponding term is treated
as perturbation. But beyond the perturbation theory these
equations possess fictitious unphysical solutions. Let us
demonstrate it explicitly with the simplest possible example of
free motion. Free covariant equations are (see, e.g.,~\cite{yb}):
\beq\label{free}
\frac{d}{d\tau}\,(mu_\mu -S_{\mu\nu}\dot{u}_\nu)=0,\quad
\dot{S}_{\mu\nu} + (u_\nu S_{\mu\lambda}-u_\mu
S_{\nu\lambda})\dot{u}_\lambda =0
\eeq
Integrating the first of them, we obtain $\;\;mu_\mu
-S_{\mu\nu}\dot{u}_\nu= c_\mu,$ where $c_\mu$ is a constant
4-vector. Then the second equation reduces to $\;\dot{S}_{\mu\nu}
= c_\mu u_\nu - c_\nu u_\mu\;$. Of course, the physical, free
solution $\;u_\mu=c_\mu/m,\; \dot{S}_{\mu\nu}=0\; $ does exist.

But equations (\ref{free}) have one more family of solutions. To
investigate and describe them, it is convenient to introduce the
spin 4-vector $S_\mu$ related to the tensor $S_{\mu\nu}$ as
follows: $S_{\mu\nu}=\ep_{\mu\nu\rho\tau}S_\rho u_\tau$. The
solutions we are looking for, can be chosen in such a way that
$S_\mu=(0,0,0,s),\;\;u_3=0,\;\;c_3=0$. Then the second of the
equations reduces to
\beq\label{prec}
s\ep_{\mu\nu\rho}\dot{u}_\rho= c_\mu u_\nu - c_\nu u_\mu, \quad
\mbox{or} \quad \dot{u}_\rho =\,\frac{1}{s}\,\ep_{\rho\mu\nu}c_\mu
u_\nu\,
\eeq
(from now on, in formulae related to this solution indices run
through 0,1,2). Equation $\;\;mu_\mu -S_{\mu\nu}\dot{u}_\nu=
c_\mu$ is satisfied identically with (\ref{prec}).

If the constant vector $c_\mu$ is time-like, we can choose the
reference frame in such a way that $c_\mu=(m/u_0,0,0)$ with
$u_0=$const (recall the condition $c_\mu u_\mu = m$). The energy
is conserved, and Eq.~(\ref{prec}) describes obviously the
precession of the particle velocity with respect to its spin with
the frequency $\om = m/u_0 s$ (in the proper time $\tau$).

Another option is a space-like $c_\mu$. Then choosing
$c_\mu=(0,0,-m/u_2)$ with $u_2=$const, we obtain
self-acceleration along the axis~1: $u_0 \sim \cosh g\tau$, $u_1
\sim \sinh g\tau$, $g=m/u_2 s$. One cannot but recall here the
self-acceleration of radiating electron in classical
electrodynamics.

Obviously, equations with pathological solutions cannot have a
fundamental meaning.

At last, let us demonstrate that it is the na\"{\i}ve, common
coordinate $\br$, rather than $\bx$, which should be considered as
the true coordinate of a relativistic spinning particle. Since
relations (\ref{def}), (\ref{defh}) are valid for a free particle
as well, the problem can be elucidated with a simple example of a
free particle with spin $1/2$. Here, instead of the Dirac
representation with the Hamiltonian of the standard form
$$H_D=\val \p + \beta m\,,$$
it is convenient to use the Foldy-Wouthuysen (FW) representation.
In it the Hamiltonian is
\[
H_{FW}=\beta \ep_{\bf p}, \quad \ep_{\bf p}=\sqrt{\p^2+m^2},
\]
and the 4-component wave functions $\psi_{\pm}$ of the states of
positive and negative energies reduce in fact to the 2-component
spinors $\phi_{\pm}$:
\[ \psi_+ = \left({\;\phi_+
\atop 0}\right), \quad \psi_- = \left({0 \atop \;\phi_-}\right)\,.
\]
Obviously, in this representation the operator of coordinate
$\hat{\br}$ defined by the usual relation \beq\label{rrr}
\hat{\br}\psi(\br)=\br\psi(\br), \eeq is just $\br$.

The transition from the exact Dirac equation in an external field
to its approximate form containing only the first-order
correction in $c^{-2}$, is performed just by means of the FW
transformation. Thus, in the resulting $c^{-2}$ Hamiltonian the
coordinate of a spinning electron is the same $\br$ as in the
completely nonrelativistic case. Nobody makes substitution
(\ref{def}) in the Coulomb potential when treating the spin-orbit
interaction in the hydrogen atom.

One more limiting case, which is of a special interest to us, is
a classical spinning particle. Such a particle is in fact a
well-localized wave packet constructed from positive-energy
states, i.e., it is naturally described in the FW representation.
Therefore, it is just $\br$ which it is natural to consider as the
coordinate of a relativistic spinning particle.

A certain subtlety here is that in the Dirac representation the
operator $\hat{\br}$ is nondiagonal. However, the operator
equations of motion certainly have the same form both in the
Dirac and Foldy-Wouthuysen representations. Correspondingly, the
semiclassical approximation to both is the same. In particular,
the time derivatives in the left-hand side of classical equations
of motion are taken of the same coordinate $\br$, which serves as
an argument of the fields in the right-hand side of these
equations.

\section{Effects of higher order in spin.\\ The idea of general formalism}

The effects linear in spin for the motion of a spinning particle
in an electromagnetic field are described by the noncovarant
Hamiltonian (\ref{lse}). The noncovarant Hamiltonian for
first-order spin effects in a gravitational field can be also
obtained from (\ref{lse}) (see~\cite{khp1,khps}) by putting $g=2$
and substituting
\begin{equation}\label{cornon}
\frac{e}{m}B_i \longrightarrow
-\,\frac{1}{2}\varepsilon_{ikl}\gamma_{klc}u^c; \;\;\;
\frac{e}{m}E_i \longrightarrow \gamma_{0ic}u^c,\;\;\; \gamma
=\,{1 \over \sqrt{1-v^2}}\,\longrightarrow u^0_w.
\end{equation}
Here $
\gamma_{abc}=\,e_{a\mu;\nu}e^\mu_{b}e^\nu_{c}=\,-\gamma_{bac}$
are the Ricci rotation coefficients. A subscript $w$ is attached
to the quantity $u^0_w$ to emphasize that $u^0_w$ is a world, but
not a tetrad, component of 4-velocity. All other indices in
expression (\ref{cornon}) are tetrad ones,
$a,b,c=0,1,2,3;\;\;i,k,l=1,2,3$.

However, at least in the motion of rotating black holes (and
possibly in some subtle spin effects for polarized nuclei of high
spin in storage rings) the interaction of second order in spin
may manifest itself. Anyway, going beyond the linear
approximation in spin is of a certain theoretical interest. To
study this general problem, a more sophisticated approach is
needed. It is based on the following physically obvious argument:
as long as we do not consider excitations of internal degrees of
freedom of a body moving in an external field, this body (even if
it is a macroscopic one!) can be treated as an elementary
particle with spin. Thus, the Hamiltonian of the spin interaction
with an external field can be derived from the elastic scattering
amplitude of a particle with spin $s$ by external field. In this
way we can describe the interaction of a relativistic particle
to arbitrary order in the spin.

The details of the approach can be found in~\cite{khp1,khps}.
Here we present only the expression for the second-order (in
spin) electromagnetic interaction:
\[ H_{e2}=-\,{Q \over 2s(2s-1)}\,\left[(\s\Na)\,-\,
{\gamma \over \gamma+1}\, (\bv\s)(\bv \Na) \right] \]
\[ \times \left[(\s\E)\,
-\,{\gamma \over \gamma+1}\,(\s\bv)(\bv\E)\,
+\,(\s[\bv\times\B])\right] \]
\begin{equation}\label{e2}
+\,{e \over 2m^2}\,{\gamma \over \gamma+1} \,\left(\s\,[\bv\times
\Na]\right) \left[\left(g-1+\,{1 \over \gamma}\right)(\s\B)\right.
\end{equation}
\[ \left. \,-\,
(g-1)\,{\gamma \over \gamma+1}\,(\s\bv)(\bv\B) -\,\left(g-{\gamma
\over \gamma+1}\right) \left(\s\,[\bv\times \E]\right)\right]. \]
Here the particle quadrupole moment $Q$ is defined as usual:
$Q\,=\,Q_{zz}\vert_{s_z=s}$.

Of great interest is the asymptotic behaviour of the interaction
(\ref{e2}) at $\gamma$~$\rightarrow$~$ \infty$. Though both
$Q$-dependent and $Q$-independent parts of the interaction
(\ref{e2}) grow up when taken separately, there is a singled out
value of the quadrupole moment for which this interaction as a
whole falls down with energy.

It is well-known (and follows immediately from (\ref{lse})) that
there is a special value of the $g$-factor, $g=2$, for which the
electromagnetic interaction linear in spin decreases with
increasing energy. Thus, the choice $\,g=2\,$ for the bare
magnetic moment is a necessary (but insufficient!) condition of
unitarity and renormalizability in quantum electrodynamics. It
holds not only for the electron, but also for the charged vector
boson in the renormalizable electroweak theory.

The same situation takes place with the second-order spin
interaction in electrodynamics. There is a special value of the
quadrupole moment $Q$ at which this interaction as well decreases
with increasing energy. If we also assume $g=2$, this value is
\begin{equation}\label{Q}
Q\,=\,-\,s(2s-1)\,{e \over m^2}.
\end{equation}
Again, (\ref{Q}) is a necessary condition of unitarity and
renormalizability. And indeed, this is the value of the
quadrupole moment of the charged vector boson in the
renormalizable electroweak theory. For it $g=2,\, s=1,\,
Q=-\,e/m^2.$

\section{Gravimagnetic moment.\\ Multipoles of black holes}

For a binary star effects of second-order in spin  are of the same
order of magnitude as the spin-spin interaction. Second-order
spin effects in the equations of motion become substantial if at
least one component of a binary is close to an extreme black hole.

The equations of motion in an external gravitational field to any
order in spin can be obtained within our general approach as
well~\cite{khp1}. However, in this brief contribution we confine
ourselves to an instructive short-cut which allows one to derive
without lengthy calculations the so-called gravimagnetic
interaction~\cite{kh}, a gravitational analogue of the
$Q$-dependent terms in (\ref{e2}).

In fact, the analogy between first-order spin interactions in
electrodynamics and gravity is incomplete. While the
electromagnetic interaction depends on the field strength, which
is gauge-invariant, the gravitational one depends not on the
Riemann tensor, which is generally covariant, but on the Ricci
rotation coefficients, which are not. This is only natural: in a
flat space, spin which is at rest in an inertial frame, precesses
in a rotating frame.

In this respect, the second-order spin interaction discussed
below, the gravimagnetic one, which depends on the Riemann
tensor, is the gravitational analogue of the first-order spin
interactions in electrodynamics. Our starting point is the
observation that the canonical momentum $p_{\mu}$ enters a
relativistic wave equation for a particle in external
electromagnetic and gravitational fields through the combination
$\Pi_{\mu}=p_{\mu}-\,eA_{\mu} -\,(1 / 2)
\,\Sigma^{ab}\gamma_{ab\mu}$.  Here $\Sigma^{ab}$ are the
generators of the Lorentz group;
$\gamma_{ab\mu}=e^c_{\mu}\gamma_{abc}$. The commutation relation
\begin{equation}
[\Pi_{\mu}, \Pi_{\nu}]=\,-ie F_{\mu\nu} +\,{i \over
2}\,\Sigma^{ab}R_{ab\mu\nu}
\end{equation}
demonstrates the remarkable correspondence
\begin{equation}\label{an}
e F_{\mu\nu}\,\longleftrightarrow\,-\,{1 \over
2}\,\Sigma^{ab}R_{ab\mu\nu}.
\end{equation}
The squared form of the Dirac equation in an external
electromagnetic field prompts that for an arbitrary spin $s$ the
Hamiltonian $e/(2m)\,\Sigma^{ab}F_{ab}$ describes the magnetic
moment interaction for $g=2$. Clearly, for an arbitrary
$g$-factor this covariant magnetic moment interaction is
\begin{equation}\label{le1}
{\cal H}_{e1}= {eg \over 4m}\,F_{ab}\Sigma^{ab}.
\end{equation}
This is in fact a covariant form of $g$-dependent terms in the
Hamiltonian (\ref{lse}). It is natural to define in analogy with
the magnetic moment
$$\frac{eg}{2m}\Sigma^{ab}\,,$$
the gravimagnetic moment
$$-\,\frac{\kappa}{2m}\Sigma^{ab}\Sigma^{cd}\,.$$
Now, the correspondence (\ref{an}) prompts the following
gravitational analogue of the Lagrangian (\ref{le1}):
\begin{equation}\label{gm}
{\cal
H}_{gm}=\,-\,\frac{\kappa}{8m}\Sigma^{ab}\Sigma^{cd}R_{abcd}.
\end{equation}
This is what we call the gravimagnetic interaction. Let us note
that in the classical limit $\Sigma^{ab} \to
S^{ab}=\varepsilon^{abcd}S_c u_d$.

The gravimagnetic ratio $\kappa$, like the gyromagnetic ratio $g$
in electrodynamics, may have in general any value. Still, it is
natural that in gravity the value $\kappa=1$ is as singled out as
$g=2$ in electrodynamics. Indeed, the analysis of the complete
noncovariant Hamiltonian for the gravitational interaction of
second order in spin, including of course $\kappa$-independent
terms which correspond to the $Q$-independent terms in (\ref{e2}),
demonstrate that just for $\kappa=1$ this total interaction
asymptotically tends to zero with increasing energy. However, the
gravitational interaction for any spin is not renormalizable even
at $\kappa=1$.

In any case, for $g=2$ and $\kappa=1$ the equations of motion have
the simplest form. Moreover, just this value of the gravimagnetic
ratio, $\kappa=1$, follows from the wave equations for the
graviton and spin-3/2 particle in an external gravitational field.

Wave equations for particles of arbitrary spins in an external
gravitational field were previously considered in~\cite{chd}. The
equation for integer spins proposed in~\cite{chd} corresponds
also to the gravimagnetic ratio $\kappa=1$. However, the value of
$\kappa$ prescribed in~\cite{chd} for half-integer spins is
different. Even in the classical limit $s\rightarrow\infty$, this
value does not tend to unity. This obviously does not comply with
the correspondence principle: at least in this classical limit
there should be no difference between integer and half-integer
spins.

Let us come back from elementary particles to macroscopic bodies.
For a classical object the values of both parameters $g$ and
$\kappa$ depend in general on the various properties of the body.
However, for black holes the situation is different. The
gyromagnetic ratio of a charged rotating black hole is universal
(and equal to that of the electron!): $g=2$~\cite{cart}. As
universal is the gravimagnetic ratio of the Kerr black hole:
$\kappa=1$. Moreover, the electric quadrupole moment of a charged
Kerr hole also equals $Q\,=\,-\,2\, es^2 / m^2 $, the value, at
which the interaction quadratic in spin decreases with energy
(this is the obvious limit of the general formula (\ref{Q}) at
$s\rightarrow\infty$). Other, higher multipoles of a charged Kerr
hole, both electromagnetic and gravitational, as well possess
just those values which guarantee that the interaction of any
order in spin (but linear in an external field) asymptotically
decreases with increasing energy~\cite{pom}.

\bigskip

\noindent{\bf Acknowledgements.} I am grateful to A.A. Pomeransky
and R.A. Sen'kov, the review is based essentially on the results
obtained in collaboration with them. The work was supported by the
Grant for Leading Scientific Schools No.~00~15~96811, and by the
Federal Program Integration-1998 through Project No.~274.

\end{document}